\documentclass{ws-mpla}

\begin{document}

\markboth{Hyun Seok Yang}
{Emergent Geometry and Quantum Gravity}

\catchline{}{}{}{}{}

\title{Emergent Geometry and Quantum Gravity}

\author{Hyun Seok Yang}

\address{Institute for the Early Universe, Ewha Womans University, Seoul 120-750, Korea \\
hsyang@ewha.ac.kr}



\maketitle


\begin{abstract}

We explain how quantum gravity can be defined by quantizing spacetime itself.
A pinpoint is that the gravitational constant $G = L_P^2$ whose physical dimension is
of $({\rm length})^2$ in natural unit introduces a symplectic structure of spacetime
which causes a noncommutative spacetime at the Planck scale $L_P$.
The symplectic structure of spacetime $M$ leads to an isomorphism between symplectic
geometry $(M, \omega)$ and Riemannian geometry $(M, g)$ where the deformations of
symplectic structure $\omega$ in terms of electromagnetic fields $F=dA$ are transformed
into those of Riemannian metric $g$. This approach for quantum gravity allows a background
independent formulation where spacetime as well as matter fields is equally emergent
from a universal vacuum of quantum gravity which is thus dubbed
as the quantum equivalence principle.


\end{abstract}

\ccode{PACS Nos.: 11.10.Nx, 02.40.Gh, 11.25.Tq}

\section{Duality from Quantization}

What is quantum gravity ? This question would be one of the most difficult questions
we have ever faced. Quantum gravity naively means to ``quantize"
a Riemannian manifold since, according to the general relativity, gravity is the dynamics of
spacetime geometry where spacetime is a Riemannian manifold and the gravitational field
is represented by a Riemannian metric. But we are still vague how to ``quantize"
the Riemannian manifold.

Mathematically, in order to define the quantization of a dynamical system,
it is necessary to first specify an underlying Poisson structure of the dynamical system.\cite{mechanics}
The dynamical system will be described by a Poisson manifold $(M, \theta)$
where $M$ is a differentiable manifold whose local coordinates are denoted by $x^A \;
(A=1, \cdots, N={\rm dim} (M))$ and the Poisson structure
\begin{equation} \label{poisson-st}
\theta = \frac{1}{2} \sum_{A,B=1}^{N} \theta^{AB}(x)  \frac{\partial}{\partial x^A} \wedge
\frac{\partial}{\partial x^B} \in \Gamma({\wedge}^2 TM)
\end{equation}
is a (not necessarily nondegenerate) bivector field. The Poisson structure
(\ref{poisson-st}) defines an $\mathbf{R}$-bilinear antisymmetric operation
$\{ - , - \}_\theta :C^\infty(M) \times C^\infty(M) \to
C^\infty(M)$ by
\begin{equation} \label{poisson-bracket}
(f,g) \mapsto \{f,g\}_\theta = \langle \theta, df
\otimes dg \rangle = \theta^{AB}(x) \frac{\partial f(x)}{\partial x^A}
\frac{\partial g(x)}{\partial x^B}
\end{equation}
such that the bracket, called the Poisson bracket, satisfies
\begin{eqnarray} \label{p-leibniz}
&&  1) \; {\rm Leibniz \; rule}: \{f,gh\}_\theta = g
\{f,h\}_\theta + \{f,g\}_\theta h, \\
\label{p-jacobi}
&& 2) \; {\rm Jacobi \; identity}: \{f,\{g,h\}_\theta\}_\theta +
\{g,\{h,f\}_\theta\}_\theta + \{h,\{f,g\}_\theta\}_\theta = 0,
\end{eqnarray}
$\forall f,g,h \in C^\infty(M)$. A Poisson manifold appears as a
natural generalization of symplectic manifolds where the Poisson
structure $\theta$ reduces to a symplectic structure if $\theta$ is nongenerate.\cite{mechanics}

Formally, the quantization, especially the canonical quantization where
$\theta^{AB}$ in Eq.(\ref{poisson-st}) is a nondegenerate constant matrix, can be done by associating
to a commutative algebra $C^\infty(M)$ of smooth functions, a noncommutative algebra
${\cal A}_\theta$ of linear operators acting on a suitable Hilbert space ${\cal H}$. That is, the
dynamical variables $f, g \in C^\infty(M)$  in a classical system are replaced by
self-adjoint operators $\widehat{f}, \widehat{g} \in {\cal A}_\theta$
acting on ${\cal H}$ and the Poisson bracket (\ref{poisson-bracket}) is replaced
by a quantum bracket
\begin{equation} \label{qm-bracket}
\{f, g\}_\theta \; \rightarrow \; -i [\widehat{f},
\widehat{g}].
\end{equation}
This completes the quantization of the dynamical system
whose phase space $M$ now becomes noncommutative, i.e.
\begin{equation} \label{nc-space}
[\widehat{x}^A, \widehat{x}^B] = i \theta^{AB}.
\end{equation}

Note that the detailed structure of Poisson manifold $(M, \theta)$
depends on what kind of dynamical system we consider.
A prominent example is the mechanical system of classical particles
where $M$ is the particle phase space with coordinates $(x^i, p_i)$ and the Poisson structure
$\theta = \hbar \sum_i \frac{\partial}{\partial x^i} \wedge \frac{\partial}{\partial p_i}$.
Here, we intentionally inserted the Planck constant $\hbar$ into $\theta$ to emphasize that
the deformation (quantization) parameter $\hbar$ carries the physical dimension of
length times momentum, i.e., $(x \times p)$ so that $\theta$ is dimensionless.
In this case, the quantization (\ref{qm-bracket}) defines quantum mechanics
as we know very well and the particle phase space $M$ is now noncommutative, i.e.
\begin{equation} \label{nc-particle}
[x^i, p_j] = i\hbar \delta^i_j.
\end{equation}

A classical field theory is a generalization of finite-dimensional particle system
to an infinite-dimensional system where particles are described
by several continuous functions $\phi^a(\mathbf{x}, t) \in C^\infty(\mathbf{R}^{3,1})$
defined on spacetime $\mathbf{R}^{3,1}$ and their conjugate variables $\pi^a(\mathbf{x}, t)
\in C^\infty(\mathbf{R}^{3,1})$ where the index $a=1,\cdots,n$ denotes internal
degrees of freedom. In this case, the corresponding Poisson structure is defined by
\begin{equation} \label{field-poisson}
\theta = \hbar \sum_{a=1}^n \int_\Sigma \frac{\delta}{\delta \phi^a(\mathbf{x})} \wedge
\frac{\delta}{\delta \pi^a(\mathbf{x})}
\end{equation}
where $\Sigma = \mathbf{R}^3 $ is a spacelike hypersurface in spacetime.
The Poisson structure (\ref{field-poisson}) generalizes the Poisson bracket
(\ref{poisson-bracket}) to an infinite-dimensional Poisson manifold
$(\mathfrak{P}, \theta)$ as follows
\begin{equation} \label{field-bracket}
\{\Phi, \Psi\}_\theta =  \hbar \sum_{a=1}^n \int_\Sigma
\Big(\frac{\delta \Phi}{\delta \phi^a(\mathbf{x})}
\frac{\delta \Psi}{\delta \pi^a(\mathbf{x})} - \frac{\delta \Phi}{\delta \pi^a(\mathbf{x})}
\frac{\delta \Psi}{\delta \phi^a(\mathbf{x})} \Big)
\end{equation}
for any functionals $(\Phi, \Psi)$ depending on the functions $(\phi^a(\mathbf{x}),
\pi^a(\mathbf{x})) \in \mathfrak{P}$ on $\Sigma$. The canonical quantization
(\ref{qm-bracket}) for the Poisson bracket (\ref{field-bracket}) leads to quantized fields,
e.g.,
\begin{equation} \label{field-commutator}
[\phi^a(\mathbf{x}), \pi^b(\mathbf{y})] = i \hbar \delta^{ab} \delta^3
(\mathbf{x}-\mathbf{y}).
\end{equation}
Quantum field theory is therefore defined by quantizing an infinite-dimensional
Poisson manifold $(\mathfrak{P}, \theta)$ in terms of $\hbar$ again, as we clearly know.

Now consider to ``quantize" gravity. First we have to carefully contemplate about
what is the dynamical system for gravity we want to quantize.
To lift the veil, it is necessary to clearly pin down what is the underlying Poisson
manifold $(M, \theta)$ for quantum gravity. Note that gravity describes the dynamics
of spacetime geometry and it is characterized
by its own intrinsic scale given by the Newton constant $G = L_P^2$
where classical gravity corresponds to $G \to 0$ limit.\footnote{Nevertheless,
gravitational phenomena are ubiquitous in our everyday life.
The reason is that the gravitational force is only attractive and so always additive.
As a result, the standard gravitational parameter $GM$ for an astronomical body with mass $M$
is not small. For example, $GM_e = 4 \times 10^{14} \;{\rm m^3/s^2}$ for the
Earth where $M_e = 5.96 \times 10^{24}$ kg, which corresponds to 1 cm compared
to the Planck length $L_P = \sqrt{G} \sim 10^{-33}$ cm.} Furthermore,
since the gravitational constant $G = L_P^2$ carries the physical dimension
of $({\rm length})^2$ in natural unit, the Newton constant $G$, as will be seen soon,
actually signifies an intrinsic Poisson structure of spacetime
\begin{equation} \label{spacetime-poisson}
\theta = \frac{1}{2} \theta^{\mu\nu} (y)
\frac{\partial}{\partial y^\mu}
\wedge \frac{\partial}{\partial y^\nu} \in \Gamma(\wedge^2 TM).
\end{equation}
Therefore, it should be reasonable to ponder on the possibility that
the quantum gravity is defined by quantizing spacetime itself in terms of $G$ instead of $\hbar$.
In other words, quantum gravity may be defined by the Poisson manifold $(M, \theta)$, where
$M$ is a spacetime manifold with the Poisson structure (\ref{spacetime-poisson}).\cite{jhep,sigma}

Customarily, we have taken the same route to the quantization of gravity
as the conventional quantum field theory. To be precise, basic phase space variables
for canonical quantum gravity are defined by a spatial metric
$g_{ij}(\mathbf{x}) \in C^\infty(\Sigma)$ defined on
a spacelike hypersurface $\Sigma \approx \mathbf{R}^3$
and the canonically conjugate variable $\pi^{ij}(\mathbf{x}) \in C^\infty(\Sigma)$
together with Hamiltonian and diffeomorphism constraints. That is, the basic Poisson manifold
$(\mathfrak{P}, \theta)$ is defined by $(g_{ij}(\mathbf{x}), \pi^{ij}(\mathbf{x})) \in \mathfrak{P}$
with the Poisson structure (\ref{field-poisson}) where $a=(ij)$.
Thus, the conventional quantum gravity also intends to quantize an infinite-dimensional
{\it particle} (graviton) phase space associated with the metric field
$g_{\mu\nu}(x)$ (or its variants such as the Ashtekar variables or spin networks) of
Riemannian geometry. This quantization scheme is very different from the quantum gravity
defined with the Poisson structure (\ref{spacetime-poisson}) because
the quantization of Poisson manifold $(\mathfrak{P}, \theta)$ is to quantize a
{\it particle} (graviton) phase space $\mathfrak{P}$ in terms of $\hbar$
while the quantization of Poisson manifold $(M, \theta)$ is to quantize spacetime $M$
itself in terms of the Newton constant $G$.\cite{sigma}

Now we have to understand what is the origin of the spacetime Poisson structure
(\ref{spacetime-poisson}) and what is the relation between Poisson geometry $(M, \theta)$
and Einstein gravity or Riemannian geometry $(M, g)$. It is well-known that the union
of three ``fundamental" constants in Nature, the Planck constant
$\hbar$ in quantum mechanics, the universal velocity $c$ in
relativity, and the Newton constant $G$ in gravity, uniquely fixes
characteristic scales for quantum gravity:
\begin{eqnarray} \label{trio}
&& M_{P} = \sqrt{\frac{c \hbar}{G}} = 2.2 \times 10^{-5} \; {\rm g}, \nonumber \\
&& L_{P} = \sqrt{\frac{G \hbar}{c^3}} = 1.6 \times 10^{-33} \;{\rm cm}, \\
&& T_{P} = \sqrt{\frac{G \hbar}{c^5}} = 5.4 \times 10^{-44} \; {\rm s} \nonumber.
\end{eqnarray}
And it is believed that in the Planck scale $L_P$ spacetime is no longer commuting
but becomes noncommutative, i.e.
\begin{equation}\label{nc-spacetime}
    [y^\mu, y^\nu] = i \theta^{\mu\nu}.
\end{equation}
Note that the noncommutative spacetime (\ref{nc-spacetime}) arises
from the quantization (\ref{qm-bracket}) with the Poisson structure (\ref{spacetime-poisson})
like as in Eq.(\ref{nc-space}). In general, if spacetime $M$ supports a Poisson structure
such as (\ref{spacetime-poisson}), the algebra $C^\infty(M)$ of smooth functions
defined on the spacetime $M$ becomes a Lie algebra under the Poisson bracket\cite{mechanics}
\begin{equation}\label{st-bracket}
 \{f, g\}_\theta (y) = \theta^{\mu\nu}(y) \frac{\partial f(y)}{\partial y^\mu}
 \frac{\partial g(y)}{\partial y^\nu}, \qquad (f, g) \in C^\infty(M).
\end{equation}
In the case where $\theta^{\mu\nu}$ is a constant matrix of rank $2n$,
we can apply the same canonical quantization to the Poisson manifold $(M, \theta)$.
We can associate to a commutative algebra
$(C^\infty(M), \{-,-\}_\theta)$ of smooth functions defined
on the spacetime $M$, a noncommutative algebra ${\cal A}_\theta$ of
linear operators on a suitable Hilbert space ${\cal H}$. That is, the
smooth functions $f, g \in C^\infty(M)$ become noncommutative
operators (fields) $\widehat{f}, \widehat{g} \in {\cal A}_\theta$ acting
on ${\cal H}$ and the Poisson bracket (\ref{st-bracket}) is replaced by a
noncommutative bracket $[\widehat{f}, \widehat{g}] \in {\cal A}_\theta$.\cite{nc-review1,nc-review2}
As a result, spacetime becomes noncommutative after the quantization and
satisfies the Heisenberg algebra (\ref{nc-spacetime}).

Therefore, we understand that spacetime admits the intrinsic Poisson structure
(\ref{spacetime-poisson}) as long as spacetime at a microscopic world is noncommutative.
Now the pith and marrow of quantum gravity is to understand how to derive a Riemannian
geometry $(M,g)$ from a Poisson geometry $(M, \theta)$ of spacetime.
Because quantization in general introduces a new kind of duality between physical or
mathematical entities, the question is what kind of duality arises from the quantization
(\ref{nc-spacetime}) of spacetime itself. Recall that the quantization (\ref{nc-particle}) of
particle phase space introduces the wave-particle duality in quantum mechanics.
The wave-particle duality results from the fact that translations in the noncommutative
phase space (\ref{nc-particle}) are an inner automorphism of the algebra ${\cal A}_\hbar$,
i.e.
\begin{equation} \label{p-inner}
e^{i\eta_i \xi^i} \widehat{f}(x^i,p_i) e^{-i\eta_i \xi^i} = \widehat{f}(x^i + \hbar l^i,
p_i  - \hbar k_i)
\end{equation}
where $\xi^i = (x^i, p_i), \; \eta_i = (k_i, l^i)$ and $\widehat{f}(x^i, p_i) \in {\cal A}_\hbar$.
The infinitesimal form of (\ref{p-inner}) is given by
\begin{equation}\label{p-deriv}
[p_i, \widehat{f}] = - i \hbar \Big\langle \frac{\partial \widehat{f}}{\partial x^i} \Big\rangle,
\quad   [x^i, \widehat{f}] = i \hbar \Big\langle \frac{\partial \widehat{f}}{\partial p_i} \Big\rangle
\end{equation}
where $\langle \cdots \rangle$ indicates a symmetric Weyl ordering.
The exactly same mathematical structure will also appear in the noncommutative spacetime
(\ref{nc-spacetime}) where translations are an inner automorphism of the noncommutative
$\star$-algebra ${\cal A}_\theta$, i.e.,\cite{nc-review1,nc-review2}
\begin{equation} \label{s-inner}
e^{ik_\mu y^\mu} \widehat{f}(y) e^{-ik_\mu y^\mu} = \widehat{f}(y + \theta \cdot k)
\end{equation}
for any $\widehat{f}(y) \in {\cal A}_\theta$.
What is a duality resulting from this inner automorphism ?

First, let us expose that the noncommutative spactime (\ref{nc-spacetime}) introduces
a new kind of duality between gauge theory and gravity.\cite{rivelles,mpl,np,steinacker}
In order to illuminate the issue in a broad context,
let us return to the system (\ref{trio}) of physical constants. We believe that
all the four interactions in Nature, gravitational, electromagnetic, weak and strong forces,
will be unified into a single force at the Planck scale (\ref{trio}). So it may be more natural
to treat gauge theory on an equal footing with gravity in the system (\ref{trio})
which is missing the gauge theory counterpart. For the reason, consider the quartet of
physical constants by adding a coupling constant $e$ which is the electric charge but sometimes
it will be denoted with $g_{YM}$ to refer to a general gauge coupling constant.
Using the symbol $L$ for length, $T$ for time, $M$ for mass, and
writing $[X]$ for the dimensions of some physical quantity $X$, we
have the following in $D$ dimensions
\begin{eqnarray} \label{quartet}
&& [\hbar] = M L^2 T^{-1} , \qquad \qquad [c] = L T^{-1}, \nonumber \\
&& [e^2] = M L^{D-1} T^{-2}, \qquad [G] = M^{-1} L^{D-1} T^{-2}.
\end{eqnarray}
A remarkable point of the system (\ref{quartet}) is that it specifies
the following intrinsic scales independently of dimensions \cite{sigma}:
\begin{equation} \label{3-quartet}
M^2 = \Big[ \frac{e^2}{G} \Big], \quad
L^2 = \Big[ \frac{G \hbar^2}{e^2 c^2} \Big], \quad
T^2 = \Big[ \frac{G \hbar^2}{e^2 c^4} \Big].
\end{equation}
From the four dimensional case where $e^2/\hbar c
\approx 1/137$, we can see that the scales in (\ref{3-quartet}) are not
so different from the Planck scales in (\ref{trio}).

Note that the first relation $GM^2 = e^2$ in (\ref{3-quartet}) implies
that at the mass scale $M$ the gravitational and electromagnetic
interactions become of equal strength. Then, the length $L$ in
Eq.(\ref{3-quartet}) is the Compton wavelength of mass $M$ where the
gravitational and electromagnetic interactions have the same
strength, which turns out to be the scale of spacetime noncommutativity where
the conspiracy between gravity and gauge theory takes place.
Suppose that a gauge theory whose coupling constant is given by $e$ is defined
in the noncommutative spacetime (\ref{nc-spacetime}). In this case, the noncommutative gauge
theory bears an intrinsic length scale given by $L^2 = |\theta|$
and then the quartet system (\ref{3-quartet}) implies that the Newton constant $G$ can be
determined by field theory parameters only, i.e.\cite{jhep}
\begin{equation}\label{coupling-dual}
\frac{G\hbar^2}{c^2} \sim e^2  |\theta|,
\end{equation}
hinting an intimate correspondence between gravity and gauge theory.\footnote{The relation (\ref{coupling-dual}) would be an analogue of the de Broglie relation $\lambda =
\frac{2\pi \hbar}{p}$ for the wave-particle duality. The de Broglie relation is possible because
quantum mechanics has a conversion factor $\hbar$ with the physical dimension
of length times momentum. Likewise, if spacetime is noncommutative and so the theory
equips with a dimensionful parameter $|\theta|$ of $({\rm length})^2$, the relation
(\ref{coupling-dual}) shows that the gravitational interaction can be inherited from
a gauge field interaction, so leading to the gauge/gravity duality.}
This novel duality in noncommutative spacetime will be clarified in the next section.

\section{Symplectization of Spacetime Geometry}

We have speculated in the previous section that, if spacetime $M$ admits a Poisson structure
such as (\ref{spacetime-poisson}), there will be a novel duality between Riemannian geometry
$(M, g)$ and Poisson geometry $(M, \theta)$. Now we will briefly sketch how this
remarkable duality can be true. We refer to Ref.~\refcite{jhep}
and a recent review\cite{sigma} for a full exposition.
See also Refs. \refcite{mpl-review,stein-review} and references therein for related discussions.

To simplify the argument,\footnote{For all mathematical details in this section,
we refer to Ref.~\refcite{mechanics}.} let us assume that the Poisson structure $\theta: T^* M
\to TM$ in (\ref{spacetime-poisson}) is nondegenerate at any point $y \in M$.
Then, we can invert this map to obtain the map $ \theta^{-1}
\equiv B: TM \to T^* M$, which is called a symplectic structure of $M$, i.e.,
a nondegenerate closed 2-form, $dB=0$, in $\Gamma(\wedge^2 T^* M)$.
The pair $(M, \omega_0 = B)$ is called a symplectic manifold.
Now consider an arbitrary deformation of the symplectic geometry $(M, B)$ by adding
an arbitrary 2-form $F \in \Gamma(\wedge^2 T^* M)$ such that $\omega_1 = B+F$.
But we will require that the resulting geometry after the deformation is still symplectic,
i.e. $dF = 0$. Then, according to the Poincar\'e lemma, the closed 2-form $F$ can locally be
written as $F=dA$ where $A \in \Gamma(T^* M)$ is an arbitrary one-form.
Because the original symplectic structure $B$ is a nondegenerate
and closed 2-form, the associated map $B: TM \to T^* M$ is a
vector bundle isomorphism. Therefore, there exists a natural pairing
$\Gamma(TM) \to \Gamma(T^* M): X \mapsto B (X) = \iota_X B$
between $C^\infty$-sections of tangent and cotangent bundles.
Note that $X \in \Gamma(TM)$ is an arbitrary vector field so that $\iota_X B$ is
an arbitrary one-form for a given $B$. Therefore, we can identify
\begin{equation}\label{moser-eq}
A= - \iota_X B
\end{equation}
and so $F=dA = - (d\iota_X + \iota_X d) B = -{\cal L}_X B$ where ${\cal L}_X
= d\iota_X + \iota_X d$ is the Lie derivative along the flow of a vector field $X$.
To conclude, the deformation of symplectic geometry $(M, B)$ in terms of an arbitrary
closed 2-form $F$ can be represented as
\begin{equation}\label{def-symp}
    \omega_1 = B- {\cal L}_X B = (1- {\cal L}_X) \omega_0.
\end{equation}

The result (\ref{def-symp}) shows that a smooth family $\omega_t =
\omega_0 + t(\omega_1 - \omega_0)$ of symplectic structures joining
$\omega_0$ to $\omega_1$ is all deformation-equivalent and there
exists a map $\phi: M \times
\mathbf{R} \to M$ as a flow -- a one-parameter family of
diffeomorphisms -- generated by the vector field $X_t$ satisfying
$\iota_{X_t} \omega_t + A = 0$ such that $\phi_t^*(\omega_t) = \omega_0$ for all $0 \leq t \leq 1$.
This can be explicitly checked by considering a local Darboux chart
$(U; y^1, \cdots, y^{2n})$ centered at $p \in U$ and valid on an open
neighborhood $U \subset M$ such that $\omega_0|_U =
\frac{1}{2} B_{\mu\nu} dy^\mu \wedge dy^\nu$ where $B_{\mu\nu}$ is a constant
symplectic matrix of rank $2n$. Now consider a flow $\phi_t: U
\times [0,1] \to M$ generated by the vector field $X_t$.
Under the action of $\phi_\epsilon$ with an
infinitesimal $\epsilon$, we find that a point $p \in U$ whose
coordinate is $y^\mu$ is mapped to $\phi_\epsilon (y) \equiv x^\mu(y) =
y^\mu + \epsilon X^\mu(y)$. Using the inverse map $\phi^{-1}_\epsilon:
x^\mu \mapsto y^\mu (x)= x^\mu - \epsilon X^\mu(x)$, the symplectic
structure $\omega_0|_U = \frac{1}{2} B_{\mu\nu} (y) dy^\mu \wedge dy^\nu$
can be expressed as\cite{hsy-siva}
\begin{eqnarray} \label{moser}
(\phi^{-1}_\epsilon)^* (\omega_0|_y) &=& \frac{1}{2} B_{\mu\nu}(x -
\epsilon X) d(x^\mu - \epsilon X^\mu)
\wedge d(x^\nu - \epsilon X^\nu) \nonumber \\ & \approx & \frac{1}{2} \Big[ B_{\mu\nu} -
\epsilon X^\lambda (\partial_\lambda B_{\mu\nu} + \partial_\nu B_{\lambda \mu}
+ \partial_\mu B_{\nu \lambda}) \nonumber \\
&& + \epsilon \Big( \partial_\mu (B_{\nu\lambda} X^\lambda)
- \partial_\nu (B_{\mu\lambda} X^\lambda) \Big) \Big] dx^\mu \wedge dx^\nu
\nonumber \\ &\equiv& B + \epsilon F
\end{eqnarray}
where $A_\mu(x) = B_{\mu\nu}(x) X^\nu(x)$ or $\iota_X B + A = 0$ and $dB
= 0$ was used for the vanishing of the second term. Equation (\ref{moser})
can be rewritten as $\phi_\epsilon^* (B + \epsilon F) = B$, which is
exactly the result obtained from (\ref{def-symp}) by taking
$\phi_\epsilon^* = (1 + \epsilon {\cal L}_X)$.

So far, $F = dA = - {\cal L}_X B \in \Gamma(\wedge^2 T^* M)$ in Eq.(\ref{def-symp})
is an arbitrary closed 2-form deforming
the original symplectic structure $B$. Now we make an important identification that
the one-form $A \in \Gamma(T^* M)$ is a connection of $U(1)$ bundle supported on
a symplectic manifold $(M,B)$ and $F = dA$ as its curvature.\cite{seib-witt} To phrase in physics,
we are considering electromagnetic fields $F=dA$ defined on a symplectic manifold $(M,B)$.
This identification is consistent with the Bianchi identity $dF=0$ in electromagnetism.
Furthermore, the identification (\ref{moser-eq}) is defined up to symplectomorphisms,
i.e., $X \sim X + X_\phi \Leftrightarrow  A \sim A + d \phi$ where $\iota_{X_\phi} B +
d \phi = 0$ and so ${\cal L}_{X_\phi} B = 0$, because it does not affect the symplectic structure
$\omega_1$ or the curvature 2-form $F=dA$. From the gauge theory point of view,
the symplectomorphisms can be identified with $U(1)$ gauge transformations.\cite{epl,ijmp,epj}
In other words, the gauge symmetry acting on $U(1)$ gauge fields as $A \to A + d \phi$
is generated by a Hamiltonian vector field $X_\phi$, i.e.,
satisfying $\iota_{X_\phi}B + d \phi =0$.

So we arrive at an overwhelming evidence for the novel duality between
Riemannian geometry $(M, g)$ and symplectic geometry $(M, \omega)$.\cite{jhep,sigma}
The first important point is that the $U(1)$ gauge symmetry is a diffeomorphism symmetry
generated by a vector field $X_\phi$ satisfying ${\cal L}_{X_\phi} B = 0$
which is known as the symplectomorphism in symplectic geometry.
Therefore, the $U(1)$ gauge symmetry on a symplectic manifold $(M, B)$ should be regarded
as a spacetime symmetry rather than an internal symmetry.
This result implies that $U(1)$ gauge fields on a symplectic spacetime
can be realized as a spacetime geometry like as gravity in
general relativity.\cite{jhep,sigma} In general relativity, the
equivalence principle beautifully explains why the gravitational
force has to manifest itself as a spacetime geometry. If the
gauge/gravity duality is realized in noncommutative spacetime, it is necessary to realize
a corresponding equivalence principle for the geometrization of the electromagnetic force.

Note that we have already realized such a noble form of the equivalence principle
in Eq.(\ref{moser}) as follows.\cite{jhep} The presence of dynamical gauge fields on a
symplectic manifold $(M,B)$ appears only as a deformation of the
symplectic manifold $(M,B)$ such that the resulting symplectic
structure is given by $\omega_1 = B+F$ where $F = dA$.
But the result (\ref{def-symp}) immediately implies that the electromagnetic force $F = dA$
can always be eliminated by a local coordinate transformation generated by a
vector field $X$ satisfying (\ref{moser-eq}), as was explicitly shown in (\ref{moser}).
Remarkably, the Poisson structure (\ref{spacetime-poisson}) of spacetime
admits the novel equivalence principle even for the electromagnetic force
which turns out to be the crux of the gauge/gravity duality.\cite{jhep,sigma}
This geometrization of the electromagnetism is inherent as an intrinsic property in
symplectic geometry known as the Darboux theorem or the Moser lemma.\cite{mechanics}
As a consequence, the electromagnetism on a symplectic spacetime can be
realized as a geometrical property of spacetime like gravity.\footnote{It should be emphasized
that there is no need to introduce any Riemannian structure to realize
the equivalence principle for the electromagnetic force. It can be derived only in the context of symplectic or Poisson geometry using the Poincar\'e lemma and the bundle isomorphism $B:TM \to T^* M$ together with the Cartan's magic formula $\mathcal{L}_X = d \iota_X + \iota_X d$.}

One overarching lesson we have learned so far is that the Darboux theorem in
symplectic geometry manifests itself as a novel form of the
equivalence principle such that the electromagnetism on symplectic spacetime
can be regarded as a theory of gravity. Hence, the final touch for the gauge/gravity duality
is to find an explicit map between electromagnetism and gravity.\cite{epl,ijmp,epj} First, note that
the $U(1)$ gauge field (\ref{moser-eq}) deforming an underlying symplectic structure
is completely encoded into a local trivialization of the symplectic structure
up to symplectomorphisms via the Darboux theorem or the Moser lemma.\cite{corna,schupp}
Let us denote the local coordinate transformation $\phi: y^\mu \mapsto x^\mu(y)$ as
\begin{equation}\label{darboux-coord}
 x^\mu(y) \equiv y^\mu + \theta^{\mu\nu} \widehat{A}_\nu (y) \in C^\infty(M),
\end{equation}
where the local coordinates $\{y^\mu\}$ will be assumed to be Darboux coordinates so that
the Poisson bracket $\{y^\mu, y^\nu\}_\theta = \theta^{\mu\nu}$ is a constant matrix
of rank $2n$. The argument (\ref{moser}) shows that such Darboux coordinates always exist.
It is well-known\cite{mechanics} that, for a given Poisson algebra $(C^\infty(M),
\{-,-\}_\theta)$, there exists a natural map $C^\infty (M) \to
TM: f \mapsto X_f$ between smooth functions in $C^\infty(M)$ and
vector fields in $TM$ such that
\begin{equation} \label{vec-map}
X_f(g) (y) \equiv \{g, f\}_\theta (y) = \Big(\theta^{\mu\nu} \frac{\partial f(y)}{\partial y^\nu}
\frac{\partial}{\partial y^\mu} \Big)  g(y)
\end{equation}
for any $g \in C^\infty(M)$. That is, we can obtain a vector field $X_f = X_f^\mu \partial_\mu
\in \Gamma(TM_y)$ from a smooth function $f \in C^\infty(M)$ defined at $y \in M$
where $X^\mu_f(y) = \theta^{\mu\nu} \frac{\partial f(y)}{\partial y^\nu}$.
As long as $\theta \in \Gamma(\wedge^2 TM)$ in (\ref{vec-map})
is a Poisson structure of $M$, the assignment (\ref{vec-map}) between a Hamiltonian
function $f$ and the corresponding Hamiltonian vector field $X_f$ is the Lie algebra
homomorphism in the sense
\begin{equation} \label{poisson-lie}
X_{\{f,g\}_{\theta}} = - [X_f,X_g]
\end{equation}
where the right-hand side represents the Lie bracket between the
Hamiltonian vector fields.

From the above arguments, we see that $U(1)$ gauge fields on a symplectic manifold
$(M,B=\theta^{-1})$ have been transformed into
a set of smooth functions
\begin{equation}\label{set-cc}
\{D_\mu(y) \in C^\infty(M)|D_\mu(y) \equiv B_{\mu\nu} x^\nu(y) = B_{\mu\nu} y^\nu
+ \widehat{A}_\mu(y), \; \mu,\nu = 1, \cdots, 2n\}
\end{equation}
through the coordinate transformation (\ref{darboux-coord}).
Thus, we can apply the map (\ref{vec-map}) to embody the Lie algebra
homomorphism (\ref{poisson-lie}) from the Poisson algebra $(C^\infty(M), \{-,-\}_\theta)$
for the set (\ref{set-cc}) to the Lie algebra $(\Gamma(TM), [-,-])$ of vector fields defined by
\begin{equation}\label{lie-vector}
\{V_\mu = V_\mu^a \partial_a \in \Gamma(TM)|V_\mu(f)(y) \equiv \{D_\mu(y), f(y)\}_\theta,
 \;a = 1, \cdots, 2n\}
\end{equation}
for any $f \in C^\infty(M)$. The vector fields $V_\mu = V_\mu^a
(y) \frac{\partial}{\partial y^a} \in \Gamma(TM_y)$  in (\ref{lie-vector}) take values
in the Lie algebra of volume-preserving diffeomorphisms because
$\partial_a V_\mu^a = 0$ by definition. But it can be shown
that the vector fields $V_\mu \in \Gamma(TM)$ are related to the
orthonormal frames (vielbeins) $E_\mu$ by $V_\mu =
\lambda E_\mu$ where $\lambda^2 = \det V_\mu^a$.
Therefore, we see that the Darboux theorem in symplectic geometry
implements a deep principle to realize a Riemannian manifold as an
emergent geometry from symplectic gauge fields through the correspondence
(\ref{vec-map}) whose metric is given by
\begin{equation} \label{emergent-metric}
ds^2 = \delta_{\mu\nu} E^\mu \otimes E^\nu = \lambda^2 \delta_{\mu\nu} V^\mu_a V^\nu_b
dy^a \otimes dy^b
\end{equation}
where $E^\mu = \lambda V^\mu \in \Gamma(T^*M)$ are dual one-forms.\cite{jhep,sigma}

Note that the emergent metric (\ref{emergent-metric}) is completely determined by
the set (\ref{set-cc}) of $U(1)$ gauge fields and it describes any Riemannian manifold
with a fixed asymptotic background. For example, the asymptotic background geometry
is a flat Euclidean space $\mathbf{R}^4$ if $\widehat{A}_\mu (y) \; (\mu = 1,\cdots, 4)$
in (\ref{set-cc}) are asymptotically vanishing fluctuations in four dimensions
while it becomes a hyper-K\"ahler manifold if $\widehat{A}_\mu (y) \; (\mu = 1,\cdots, 4)$
describe a noncommutative $U(1)$ instanton.\cite{pl,prl}

\section{Quantization of Gravity}

Let us recapitulate how we could get the Riemannian metric (\ref{emergent-metric}).
We have considered electromagnetism on a symplectic spacetime $(M, B)$. The electromagnetic
fields in this case manifest themselves only as a deformation of
symplectic structure such that the resulting symplectic spacetime is described
by $(M, B+F)$ where $F=dA$. Via the Darboux theorem together with the homomorphism
(\ref{vec-map}), this deformation of symplectic structure in terms of
the electromagnetic force  $F=dA$ can be translated into a deformation of frame bundle
over spacetime manifold $M$; $\partial_\mu \to E_\mu = E^a_\mu(y) \partial_a$,
or, in terms of dual frames, $d y^\mu \to E^\mu = E_a^\mu(y) dy^a$.
That is, the deformations of symplectic spacetime $(M,B)$ in terms of electromagnetic
force $F=dA$ are isomorphic to those of Riemannian manifold $(M,g)$
\begin{equation}\label{def-riemann}
    ds^2 = \delta_{\mu\nu} dy^\mu \otimes dy^\nu \; \to \;
    ds^2 = \delta_{\mu\nu} E^\mu \otimes E^\nu.
\end{equation}
This isomorphism implies that a field theory equipped with the fields in (\ref{set-cc})
on a symplectic or Poisson spacetime gives rise to Einstein gravity.\cite{jhep,sigma}
Another crucial point, as will be shown below, is that an underlying field theory action
for emergent gravity will be represented only by the Poisson bracket $\{D_\mu, D_\nu\}_\theta (y)$
between the fields in (\ref{set-cc}), and so the equations of motion will be defined
only with the Poisson bracket (\ref{st-bracket}).
If this is the case, quantum gravity will be much more accessible
since there is a natural symplectic or Poisson structure (\ref{spacetime-poisson})
and so it is obvious how to quantize the underlying system,
as was already done in (\ref{qm-bracket}).

We demonstrate the emergent gravity with the following action
\begin{equation} \label{symp-action}
S_P =  \frac{1}{4g_{YM}^2} \int d^{2n} y  \{D_\mu(y),
D_\nu(y)\}_\theta \{D^\mu(y), D^\nu(y)\}_\theta
\end{equation}
where  $g_{YM}$ is a $2n$-dimensional gauge coupling constant.
Note that
\begin{eqnarray} \label{poisson-dd}
\{D_\mu (y),D_\nu (y)\}_\theta &=& - B_{\mu\nu} + \partial_\mu
\widehat{A}_\nu(y) - \partial_\nu \widehat{A}_\mu (y)
+ \{\widehat{A}_\mu (y), \widehat{A}_\nu (y)\}_\theta \nonumber \\ &\equiv&
- B_{\mu\nu} + \widehat{F}_{\mu\nu}(y)
\end{eqnarray}
and
\begin{eqnarray} \label{poisson-ddd}
\{D_\mu(y), \{D_\nu (y),D_\lambda (y)\}_\theta \}_\theta &=&
\partial_\mu \widehat{F}_{\nu\lambda}(y) +
\{\widehat{A}_\mu (y), \widehat{F}_{\nu\lambda} (y)\}_\theta \nonumber \\
&\equiv& \widehat{D}_\mu \widehat{F}_{\nu\lambda}(y).
\end{eqnarray}
It is easy to see by identifying $f(y) = D_\mu(y)$ and $g(y) = D_\nu(y)$ and
using the relation (\ref{poisson-dd}) that the Lie algebra homomorphism
(\ref{poisson-lie}) leads to the following identity
\begin{equation} \label{map-fvv}
X_{\widehat{F}_{\mu\nu}} = [V_\mu, V_\nu]
\end{equation}
where $V_\mu \equiv X_{D_\mu}$ and $V_\nu \equiv X_{D_\nu}$. Similarly,
using (\ref{poisson-ddd}), we can further deduce that
\begin{equation} \label{map-vvv}
X_{\widehat{D}_\mu \widehat{F}_{\nu\lambda}} = [V_\mu, [V_\nu, V_\lambda]].
\end{equation}

The Jacobi identity (\ref{p-jacobi}) for the Poisson bracket
(\ref{poisson-ddd}) can be written as in the form
\begin{equation} \label{poisson-jacobi}
\{D_{[\mu}, \{D_\nu, D_{\lambda]} \}_{\theta} \}_{\theta}
= \widehat{D}_{[\mu} \widehat{F}_{\nu\lambda]} = 0,
\end{equation}
where $[\mu,\nu,\lambda]$ denotes the cyclic permutation of indices.
Similarly the equations of motion derived from the action (\ref{symp-action}) read as
\begin{equation} \label{poisson-eom}
\{D^\mu, \{D_\mu, D_\nu \}_{\theta} \}_{\theta} = \widehat{D}^\mu \widehat{F}_{\mu\nu} = 0.
\end{equation}
Then, the map (\ref{map-vvv}) translates the Jacobi identity
(\ref{poisson-jacobi}) and the equations of motion (\ref{poisson-eom}) into some
geometric relations between the vector fields $V_\mu$
defined by (\ref{lie-vector}). That is, we have the following
correspondence\cite{jhep}
\begin{eqnarray} \label{map-jacobi}
&& \widehat{D}_{[\mu} \widehat{F}_{\nu\lambda]} = 0 \;\;
\Leftrightarrow \;\; [V_{[\mu}, [V_\nu, V_{\lambda]}]] = 0, \\
\label{map-eom}
&& \widehat{D}^\mu \widehat{F}_{\mu\nu} = 0 \;\;
\Leftrightarrow \;\; [V^\mu, [V_\mu, V_\nu]] = 0.
\end{eqnarray}
Since the vector fields $V_\mu$ in (\ref{lie-vector}) completely determine
the gravitational metric (\ref{emergent-metric}), the right-hand sides of (\ref{map-jacobi})
and (\ref{map-eom}) are thus second-order differential equations of
the metric (\ref{emergent-metric}), so that they finally reduce to some equations related
to Riemann curvature tensors. It was shown in Ref.~\refcite{jhep} that
the Bianchi identity (\ref{map-jacobi}) for symplectic gauge fields in the action
(\ref{symp-action}) is equal to the first Bianchi identity of Riemann tensors, i.e.
\begin{equation} \label{1st-bianchi}
[V_{[\mu}, [V_\nu, V_{\lambda]}]] = 0  \;\;
\Leftrightarrow \;\;  R_{[\mu\nu\lambda]\rho} = 0,
\end{equation}
and the equations of motion (\ref{map-eom}) are equivalent to the Einstein equations
for the emergent metric (\ref{emergent-metric}), i.e.
\begin{equation} \label{einstein-eq}
[V^\mu, [V_\mu, V_\nu]] = 0 \;\;
\Leftrightarrow \;\; R_{\mu\nu} - \frac{1}{2} g_{\mu\nu} R = \frac{8\pi G}{c^4} T_{\mu\nu},
\end{equation}
where the gravitational constant $G$ is defined by (\ref{coupling-dual}).
See Ref.~\refcite{jhep} for the derivation and, especially, for a surprising content
of the energy-momentum tensor in Eq.(\ref{einstein-eq}).

Now we have realized that Einstein gravity can be emergent from electromagnetism
as long as spacetime admits a symplectic structure (\ref{spacetime-poisson}).
Therefore classical gravity is defined by the action (\ref{symp-action}) and
the so-called emergent gravity suggests a novel and authentic way for quantum gravity
where the quantization of gravity is reduced to quantizing a dynamical system
described by the action (\ref{symp-action}).\cite{sigma} With the Poisson structure (\ref{spacetime-poisson}), who is still in agony to find a quantum world ?

A question is whether the canonical quantization (\ref{qm-bracket}) for the action
(\ref{symp-action}) correctly describes quantum spacetime geometries and resolves some notorious
problems in classical gravity, e.g., the cosmological constant problem.\cite{cc-hsy1,cc-hsy2}
We will not try to answer to the question right now because it may be premature to disclose.
Instead we will show that the action (\ref{symp-action}) arises in the commutative limit
of a completely background independent theory where no prior existence of spacetime
is assumed but is defined by the theory itself as a vacuum solution.

Consider the zero-dimensional IKKT matrix model\cite{ikkt} whose action
is given by
\begin{equation} \label{ikkt}
S_{IKKT} =  - \frac{1}{4} {\rm Tr} \big( [X_\mu, X_\nu][X^\mu, X^\nu] \big).
\end{equation}
Since the action (\ref{ikkt}) is 0-dimensional, it does not assume any kind of
spacetime structure. There are only a bunch of
$N \times N$ Hermitian matrices $X^\mu \; (\mu=1, \cdots, 2n)$ which are
subject to a couple of algebraic relations given by
\begin{eqnarray} \label{matrix-bianchi}
&& [X^{[\mu}, [X^\nu, X^{\lambda]}]= 0, \\
\label{matrix-eom}
&& [X_\mu, [X^\mu, X^\nu]] = 0.
\end{eqnarray}
In order to expand the matrix theory (\ref{ikkt}), first we have to
specify a vacuum of the theory where all fluctuations are supported.
Of course, the vacuum solution should also satisfy the
Eqs. (\ref{matrix-bianchi}) and (\ref{matrix-eom}). Suppose that the
vacuum solution is given by $X^\mu_{{\rm vac}} = y^\mu$.\footnote{It should be remarked that
a sufficient condition for the vacuum is that it is a (semi-)stable solution of
the theory. Therefore the vacuum is not unique in general. For example,
an instanton solution in four dimensions, $[X^\mu_{{\rm ins}}, X^\nu_{{\rm ins}}] =
\pm \frac{1}{2} {\varepsilon^{\mu\nu}}_{\lambda\rho}
[X^\lambda_{{\rm ins}}, X^\rho_{{\rm ins}}]$, is also a stable vacuum satisfying
(\ref{matrix-eom}). But we will understand the instanton vacuum as an inhomogeneous deformation,
possibly with a topology change, from the primitive vacuum (\ref{nc-spacetime}), i.e.,
$X^\mu_{{\rm ins}} = y^\mu + \theta^{\mu\nu} \widehat{A}^{{\rm (ins)}}_\nu(y)$.\cite{jhep}}
In the limit $N \to \infty$, the noncommutative space defined by (\ref{nc-spacetime}) definitely
satisfies the equations of motion (\ref{matrix-eom}). Furthermore, in this case, the
matrix algebra $({\cal A}_N, [-,-])$ defining the action (\ref{ikkt})
can be mapped to a noncommutative $\star$-algebra $({\cal A}_\theta,
[-,-]_\star)$\cite{aiikkt,epj} defined by the star product
\begin{equation}\label{star-product}
(\widehat{f} \star \widehat{g}) (y) = \exp \Big(\frac{i}{2}\theta^{\mu\nu}
\frac{\partial}{\partial y^\mu} \frac{\partial}{\partial z^\nu}\Big) f(y) g(z) \Big|_{y=z}
\end{equation}
where $\widehat{f}, \widehat{g} \in {\cal A}_\theta$ and $f, g \in C^\infty(M)$.
For example, the large $N$ matrices $X^\mu \equiv \theta^{\mu\nu}
\widehat{D}_\nu$ can be expanded around the Moyal vacuum (\ref{nc-spacetime}) as follows
\begin{equation} \label{x-exp}
\widehat{D}_\mu(y) = B_{\mu\nu} y^\nu + \widehat{A}_\mu(y) \in {\cal A}_\theta.
\end{equation}
It is then easy to calculate the following $\star$-commutator
\begin{eqnarray} \label{nc-curvature}
-i [\widehat{D}_\mu (y), \widehat{D}_\nu (y)]_\star &=& -B_{\mu\nu} + \partial_\mu
\widehat{A}_\nu(y) - \partial_\mu \widehat{A}_\nu (y)
- i [\widehat{A}_\mu (y), \widehat{A}_\nu (y)]_\star \nonumber \\ &\equiv&
 - B_{\mu\nu} + \widehat{F}^\star_{\mu\nu}(y).
\end{eqnarray}
Therefore we see that the matrix action (\ref{ikkt}) can be obtained by quantizing \`a la (\ref{qm-bracket}) the classical action (\ref{symp-action}).

The noncommutative $\star$-algebra $({\cal A}_\theta, [-,-]_\star)$ can be obtained
by the canonical quantization (\ref{qm-bracket}) of Poisson algebra
$(C^\infty(M), \{-,-\}_\theta)$ through the Weyl-Wigner correspondence\cite{nc-review1,nc-review2}
where the Poisson structure is defined by (\ref{spacetime-poisson}).
In particular, the correspondence (\ref{vec-map}) between the Poisson algebra
$(C^\infty(M), \{-,-\}_\theta)$ and vector fields in $\Gamma(TM)$ can be generalized
to the noncommutative $\star$-algebra $({\cal A}_\theta, [-,-]_\star)$ by considering
an adjoint action of noncommutative gauge fields in (\ref{x-exp}) as follows
\begin{eqnarray} \label{nc-vector}
\widehat{V}_\mu [\widehat{f}] (y) &\equiv & - i [
\widehat{D}_\mu (y), \widehat{f}(y) ]_\star = - \theta^{ab}
\frac{\partial D_\mu(y)}{\partial y^b} \frac{\partial f(y)}{\partial y^a}
+ \cdots \nonumber \\ &=& V_\mu[f](y) + {\cal O}(\theta^3).
\end{eqnarray}
Note that the leading term in (\ref{nc-vector}) precisely reproduces the usual vector fields
in (\ref{lie-vector}) and so we will refer to $\widehat{V}_\mu$ in (\ref{nc-vector}) as
generalized vector fields.
According to the correspondence (\ref{nc-vector}), the noncommutative gauge fields
in (\ref{x-exp}) are mapped to generalized vector fields
as an inner derivation in ${\cal A}_\theta$.
In particular, we have the following property generalizing the identity (\ref{map-fvv})
\begin{equation} \label{jacobi-derivation}
\widehat{X}_{\widehat{F}^\star_{\mu\nu}} = [\widehat{V}_\mu, \widehat{V}_\nu]_\star
\end{equation}
where $[\widehat{V}_\mu, \widehat{V}_\nu]_\star = [V_\mu, V_\nu] +
{\cal O}(\theta^3)$ is a generalization of Lie bracket to the generalized vector fields
in (\ref{nc-vector}). Using the maps (\ref{nc-vector}) and
(\ref{jacobi-derivation}), we can further deduce that
\begin{equation} \label{jacobi-3}
\widehat{X}_{\widehat{D}^\star_\mu \widehat{F}^\star_{\nu\lambda}} =
[\widehat{V}_\mu, [\widehat{V}_\nu, \widehat{V}_\lambda]_\star ]_\star.
\end{equation}
Using the relation (\ref{jacobi-3}), we can easily show that
the correspondence in (\ref{1st-bianchi}) and (\ref{einstein-eq})
can be generalized to the noncommutative gauge fields (\ref{x-exp}) and
the generalized vector fields defined by (\ref{nc-vector}) as follows \cite{jhep,sigma}:
\begin{eqnarray} \label{nc-jacobi}
&& \widehat{D}^\star_{[\mu} \widehat{F}^\star_{\nu\lambda]} = 0 \quad \Leftrightarrow
\quad
 [\widehat{V}_{[\mu}, [\widehat{V}_\nu,
\widehat{V}_{\lambda]}]_\star ]_\star = 0, \\
\label{nc-eom}
&&  \widehat{D}^{\star \mu} \widehat{F}^\star_{\mu\nu} = 0 \quad \Leftrightarrow
\quad [\widehat{V}^\mu, [\widehat{V}_\mu, \widehat{V}_\nu]_\star ]_\star = 0.
\end{eqnarray}

Since the leading order in (\ref{nc-vector}) recovers usual vector fields,
the Einstein gravity described by (\ref{1st-bianchi}) and (\ref{einstein-eq}) will appear
as the leading order of noncommutative gauge fields described by (\ref{nc-jacobi})
and (\ref{nc-eom}). Therefore we expect, according to the line of thought in Sec. 1, that
Eqs. (\ref{nc-jacobi}) and (\ref{nc-eom}) will describe quantum gravity, which can be formulated
in terms of the background independent matrix action (\ref{ikkt}). Our scheme for quantum
gravity is radically different from the conventional wisdom.\cite{sigma}

\section{Geometry and Matters from Algebra}

We have shown that the spacetime Poisson structure (\ref{spacetime-poisson}) provides
a background independent completion of quantum gravity through the IKKT matrix model (\ref{ikkt}).
We will further enhance this picture by showing that the AdS/CFT correspondence\cite{ads-cft1,ads-cft2,ads-cft3} can be understood as the emergent gravity
defined by the spacetime Poisson structure (\ref{spacetime-poisson}).

Let us consider $U(N  \to \infty)$ Yang-Mills theory in $d$
dimensions
\begin{equation} \label{n=4}
S_N = - \frac{1}{G_s} \int d^d z {\rm Tr}
\left(\frac{1}{4} F_{\mu\nu} F^{\mu\nu}  +
\frac{1}{2} D_\mu \Phi^a D^\mu \Phi^a - \frac{1}{4} [\Phi^a, \Phi^b]^2
\right),
\end{equation}
where $G_s \equiv 2\pi g_s/(2\pi\kappa)^{\frac{4-d}{2}}$ and
$\Phi^a \; (a = 1, \cdots, 2n)$ are adjoint scalar fields in $U(N)$.
Note that, if $d=4$ and $n=3$, the action (\ref{n=4}) is exactly the bosonic part of
4-dimensional ${\cal N} =4$ supersymmetric $U(N)$ Yang-Mills theory,
which is the large $N$ gauge theory of the AdS/CFT correspondence.\cite{ads-cft1,ads-cft2,ads-cft3}
Suppose that a vacuum of the theory (\ref{n=4}) is given by
\begin{equation} \label{matrix-vacuum}
\langle \Phi^a \rangle_{\rm vac} = \frac{1}{\kappa} y^a, \quad
\langle A_\mu \rangle_{\rm vac} = 0.
\end{equation}
Assume that the vacuum expectation values $y^a \in U(N \to \infty)$
satisfy the algebra
\begin{equation} \label{vacuum-algebra}
[y^a, y^b] = i \theta^{ab} \mathbf{1}_{N \times N},
\end{equation}
where $\theta^{ab}$ is a constant matrix of rank $2n$. If so, the
vacuum (\ref{matrix-vacuum}) in the $N \to \infty$ limit is definitely a solution of
the theory (\ref{n=4}) and the adjoint scalar fields in vacuum
satisfy the noncommutative Moyal algebra (\ref{nc-spacetime}).
The large $N$ matrices in the action (\ref{n=4}) can then be mapped
to noncommutative fields in ${\cal A}_\theta$ like as (\ref{x-exp}).

Let us consider fluctuations $\widehat{A}_{M} (X) \equiv
(\widehat{A}_\mu, \widehat{A}_a) (z,y), \; M = 1, \cdots, d+2n$, of large $N$ matrices
in the action (\ref{n=4}) around the vacuum (\ref{matrix-vacuum})
\begin{equation} \label{mat-fluct}
\Phi^a(z,y) = \frac{1}{\kappa} \big( y^a + \theta^{ab}
\widehat{A}_b(z,y) \big), \quad
D_\mu (z,y) = \partial_\mu - i \widehat{A}_\mu (z,y),
\end{equation}
where the fluctuations are assumed to also depend on the vacuum moduli in (\ref{matrix-vacuum}).
Therefore let us introduce $D=d+2n$-dimensional coordinates $X^M = (z^\mu,
y^a)$ and $D$-dimensional connections defined by
\begin{eqnarray} \label{decomp-cov}
D_M (X) &=& \partial_M - i \widehat{A}_M(X) \nonumber \\ &\equiv& (D_\mu =
\partial_\mu - i \widehat{A}_\mu, D_a = -i\kappa B_{ab}
\Phi^b)(z,y).
\end{eqnarray}
As a result, the large $N$ matrices in the action (\ref{n=4}) are now represented
by their master fields which are higher dimensional noncommutative $U(1)$ gauge fields in
(\ref{decomp-cov}) whose field strength is given by
\begin{equation} \label{D-field}
\widehat{F}_{MN} = \partial_M \widehat{A}_N - \partial_N \widehat{A}_M
- i [\widehat{A}_M, \widehat{A}_N]_\star.
\end{equation}
In the end, the $d$-dimensional $U(N)$ Yang-Mills theory
(\ref{n=4}) has been transformed into a $D$-dimensional
noncommutative $U(1)$ gauge theory and the action (\ref{n=4}) can be recast
into the simple form\cite{epj}
\begin{equation} \label{higher-action}
\widehat{S}_B = - \frac{1}{4 g^2_{YM}} \int d^D X
(\widehat{F}_{MN} - B_{MN}) \star (\widehat{F}^{MN} -
B^{MN}).
\end{equation}

To find a gravitational metric dual to the large $N$ gauge theory (\ref{n=4}) or, equivalently,
to find an emergent metric determined by the noncommutative gauge theory (\ref{higher-action}),
first apply the adjoint operation (\ref{nc-vector})
to the $D$-dimensional noncommutative gauge fields in (\ref{decomp-cov})
after switching the index $M \to A = 1, \cdots, D = d+ 2n$:
\begin{eqnarray} \label{D-vector}
\widehat{V}_A [\widehat{f}](X) &=& [D_A, \widehat{f}]_\star (z,y)
\nonumber \\
&\equiv& V_A^M(z,y) \partial_M f(z,y) + {\cal O}(\theta^3).
\end{eqnarray}
In commutative limit, the vector fields $V_A = V_A^M \partial_M \in \Gamma(TM)$
on a $D$-dimensional manifold $M$
are given by
\begin{equation}\label{vec-D}
V_A (X) = (\partial_\mu + A_\mu^a \partial_a,  D_a^b \partial_b)
\end{equation}
or their dual basis $V^A = V^A_M dX^M \in \Gamma(T^*M)$ is given by
\begin{equation}\label{form-d}
V^A (X) = \bigl(dz^\mu,  V^a_b(dy^b- A_\mu^b dz^\mu)\bigr)
\end{equation}
where $V_a^c D_c^b = \delta_a^b$ and
\begin{equation}\label{com-vec}
A_\mu^a \equiv - \theta^{ab} \frac{\partial \widehat{A}_\mu}{\partial y^b},
\qquad  D_a^b \equiv \delta^b_a - \theta^{bc} \frac{\partial \widehat{A}_a}{\partial
y^c}.
\end{equation}
Hence the $D$-dimensional metric can be determined by the dual basis (\ref{form-d}) as\cite{epj,jhep}
\begin{eqnarray} \label{d-metric}
ds^2 &=&  \lambda^2 \eta_{AB} V^A \otimes V^B  \nonumber \\
&=& \lambda^2 \Bigl(\eta_{\mu\nu} dz^\mu dz^\nu +
\delta_{ab} V^a_c V^b_d (dy^c - \mathbf{A}^c)(dy^d - \mathbf{A}^d)
\Bigr),
\end{eqnarray}
where $\mathbf{A}^a = A_\mu^a dz^\mu$ and the conformal factor is determined by
\begin{equation}\label{D-volume}
\lambda^2 = \mathfrak{V} (V_1, \cdots, V_D)
\end{equation}
for a $D$-dimensional volume form $\mathfrak{V} = d^dz \wedge \nu$.\footnote{It can be shown\cite{epj} that the vacuum geometry (\ref{d-metric}) for the state (\ref{matrix-vacuum}) is a flat spacetime
${\bf R}^{1, D-1}$ if $\nu = dy^1 \wedge \cdots \wedge dy^{2n}$ while it is
$AdS_{d+1} \times {\bf S}^{2n-1}$ if $\nu = \frac{dy^1 \wedge \cdots \wedge dy^{2n}}{\rho^2}$
with $\rho^2 = \sum_{a=1}^{2n} y^ay^a$.}

Note that the large $N$ gauge theory (\ref{n=4}) gives rise to a series of matrix models
depending on the choice of base space ${\bf R}^{1,d-1}$, which is a nonperturbative
formulation of string or M theories. (See Ref.~\refcite{taylor} for a review and references therein.)
We see that the existence of nontrivial gauge fields $A_\mu(z)$ causes the curving of
the original flat spacetime ${\bf R}^{1,d-1}$ and so it becomes dynamical together with an entirely
emergent $2n$-dimensional space. Therefore, the large $N$ gauge theory (\ref{n=4}) almost provides
a background independent description of spacetime geometry except the original
background ${\bf R}^{1,d-1}$ whose existence was {\it a priori} assumed at the outset.
We may completely remove the spacetime ${\bf R}^{1,d-1}$ from the action (\ref{n=4}) and start
with a theory without spacetime from the beginning, like as (\ref{ikkt}),\cite{jhep,sigma} by applying the `matrix T-duality' (see Sec. VI.A in Ref.~\refcite{taylor}).

A remarkable aspect of the large $N$ gauge theory (\ref{n=4}) is that it admits a rich variety
of topological objects. Consider a stable class of time-independent solutions in
the action (\ref{n=4}) satisfying the asymptotic boundary condition (\ref{matrix-vacuum}).
In particular, the matrices $\Phi^a (\mathbf{x})$ are nondegenerate along
$\mathbf{S}^{d-1}= {\bf R}^{d-1} \cup \{\infty\}$ and so $\Phi^a$ defines
a well-defined map\cite{horava}
\begin{equation} \label{homotopy}
\Phi^a: \mathbf{S}^{d-1} \to GL(N,\mathbf{C})
\end{equation}
from $\mathbf{S}^{d-1}$ to the group of nondegenerate complex $N
\times N$ matrices. If this map represents a nontrivial class in the
$(d-1)$-th homotopy group $\pi_{d-1}(GL(N,\mathbf{C}))$, the solution
(\ref{homotopy}) will be stable under small perturbations, and the
corresponding nontrivial element of $\pi_{d-1}(GL(N,\mathbf{C}))$ represents a
topological invariant. In the stable regime where $N > \frac{d-1}{2}$, the homotopy groups of
$GL(N,\mathbf{C})$ or $U(N)$ define a generalized cohomology theory, known
as K-theory $K(X)$.\cite{k-theory}  For example, for $X = \mathbf{R}^d$,
this group is given by
\begin{equation}\label{k-theory}
K(\mathbf{R}^d) = \pi_{d-1}(GL(N, \mathbf{C})).
\end{equation}

It is well-known\cite{k-theory} that K-theory generators in (\ref{k-theory}) can be constructed in
terms of Clifford module. The construction uses the gamma matrices $\Gamma^\mu: S_+
\to S_-$ of $SO(d-1,1)$ to satisfy $\{\Gamma^\mu, \Gamma^\nu \} = 2
\eta^{\mu\nu}$ of the Lorentz group $SO(d-1,1)$.\cite{k-witten,k-review}
Let $X$ be even dimensional so that $K(X) = \mathbf{Z}$ and $S_\pm$ be two irreducible spinor
representations of $Spin(d)$ Lorentz group and a Dirac operator
$\mathcal{D}: V \times S_+ \to V \times S_-$
such that $\mathcal{D} = \Gamma^\mu \partial_\mu + \cdots$
acting on a Hilbert space $V$ as well as a spinor vector space $S_\pm$.
An explicit construction\cite{jhep,sigma} shows that the Dirac operator $\mathcal{D}$
acts on collective (coarse-grained) modes of the solution (\ref{homotopy}) satisfying
the Dirac equation
\begin{equation}\label{dirac-eq}
i \Gamma^\mu(\partial_\mu - ieA_\mu - iA_\mu^I Q^I) \chi + \cdots = 0,
\end{equation}
where the fermion $\chi^A$ carries the index $A=(\alpha a)$ with $\alpha$ the spinor index
of $Spin(d)$ and $a = 1, \cdots, n$ an internal index of an $n$-dimensional
representation $End(V)$ of a compact symmetry $G$.

To conclude, we observed that the theory (\ref{n=4}) allows topologically stable
solutions as long as the homotopy group (\ref{homotopy}) is nontrivial.
Remarkably, a matter field such as leptons and quarks simply
arises from such a stable solution and non-Abelian gauge fields
correspond to collective zero-modes of the stable localized
solution.\cite{horava} Although the solution (\ref{homotopy}) is interpreted as
particles and gauge fields ignoring its gravitational effects,
we have to recall that it is a stable excitation
over the vacuum (\ref{matrix-vacuum}) and so originally a part of spacetime
geometry according to the map (\ref{D-vector}). Consequently, we get a
remarkable picture, if any, that matter fields such as leptons and
quarks simply arise as a stable localized geometry, which is a
topological object in the defining algebra (noncommutative
$\star$-algebra) of quantum gravity.\cite{jhep,sigma} This approach for quantum gravity thus
allows a background independent formulation where spacetime
as well as matter fields is equally emergent from a universal vacuum of quantum gravity
which may be dubbed as the quantum equivalence principle.

\section*{Acknowledgments}

We sincerely thank Jungjai Lee and John J. Oh for helpful discussions and their encouragements.
The work of the author was supported by the RP-Grant 2009 of Ewha Womans University.

\end{document}